%% file: paper.tex
\documentstyle[seceq,twocolumn,epsfig]{jpsj}

\newcommand{\kakkoa}[1]{\left( #1 \right)}

\newcommand{\bm}[1]{\textbf{#1}}
\newcommand{\ket}[1]{\left| #1 \right\rangle}
\newcommand{\bra}[1]{\left\langle #1 \right|}

\renewcommand{\dag}{+}

\makeatletter
\def\lsim{\mathrel{\mathpalette\gl@align<}}
\def\gsim{\mathrel{\mathpalette\gl@align>}}
\def\gl@align#1#2{\lower.6ex\vbox{\baselineskip\z@skip\lineskip\z@
    \ialign{$\m@th#1\hfil##\hfil$\crcr#2\crcr\sim\crcr}}}
\makeatother

\def\runtitle{Magnetic Order in the Double Exchange Model
 in Infinite Dimensions}
\def\runauthor{Kentaro {\sc Nagai}, Tsutomu {\sc Momoi} and 
Kenn {\sc Kubo}}

\title{Magnetic Order in the Double Exchange Model
 in Infinite Dimensions}
\author{Kentaro {\sc Nagai}$^{1}$, Tsutomu {\sc Momoi}$^{1,2}$ and 
Kenn {\sc Kubo}$^{3}$}

\inst{$^{1}$Institute of Physics, University of Tsukuba,
Tsukuba, Ibaraki 305-8571, Japan\\
$^{2}$Lyman Laboratory of Physics, Harvard University,
Cambridge, MA 02138, USA\\
$^{3}$Department of Physics, Aoyama Gakuin University,
Setagaya, Tokyo 157-8572, Japan }

\abst{
We studied magnetic properties of the double exchange (DE) model with
$S=1/2$ localized spins at $T=0$, using exact diagonalization in the framework
of the dynamical mean field theory. Obtained phase diagram contains 
ferromagnetic, antiferromagnetic and paramagnetic phases. Comparing 
the phase diagram with that of the DE model with classical localized
spins, we found that the quantum fluctuations of localized spins partly
destabilize the ferromagnetism and expand the paramagnetic phase region. We
found that phase separations occur between the antiferromagnetic and
paramagnetic phases as well as the paramagnetic and 
ferromagnetic ones. }

\kword
{double exchange model, infinite dimensions, quantum effect, phase
  diagram, ferromagnetism, antiferromagnetism, phase separation,
  density of states}

\recdate{\hspace*{5cm}}

\begin{document}
\sloppy
\maketitle
\input{paper1.tex}

\input{paper2.tex}

\input{paper3.tex}

\input{paper4.tex}

\input{paper5.tex}

\input{acknowledge.tex}

\input{biblio.tex}
\onecolumn

\begin{figure}[ht]
  \caption{The magnetic susceptibility $\chi$ for
    $J_{\mbox{{\scriptsize H}}}S=5.0$ and $n=1.0$. (a) Wave-vector
    dependence of $\chi$. (b) Temperature dependence of the
    antiferromagnetic susceptibility ($X(\bm{q})=-1.0$). $X(\bm{q})$ is
    defined by (\ref{Xdef}) in the text. } 
\vspace*{0.5cm}
\epsfig{file=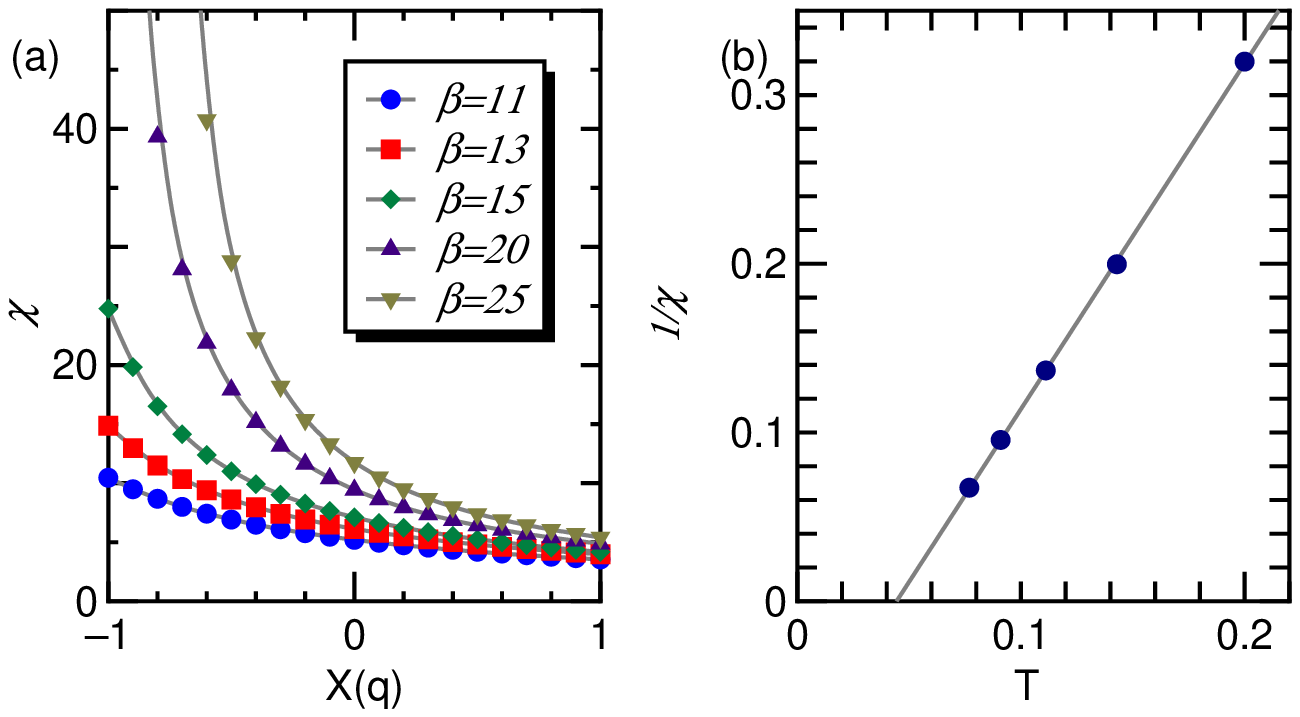}
  \label{J10susa}
\end{figure}

\begin{figure}[hb]
  \caption{The susceptibility for $J_{\mbox{{\scriptsize H}}}S=5.0$
and $n=0.8$. (a) Wave-vector dependence of the susceptibility. 
(b) Temperature dependence of the ferromagnetic 
susceptibility ($X(\bm{q})=1.0$).}
\vspace*{0.5cm}
\epsfig{file=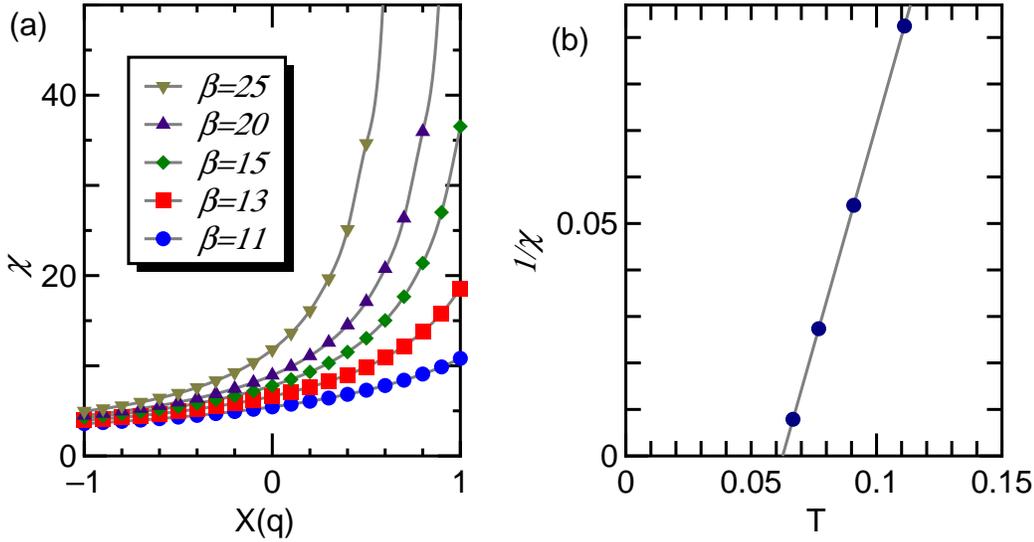}
  \label{J10susf}
\end{figure}


\begin{figure}[ht]
  \caption{Phase diagram for $S=\infty$ at $\beta=400$. Antiferromagnetic, 
    ferromagnetic and paramagnetic phases are denoted by AF, F and P, 
    respectively. PS means the region where phase separation occurs.}
\vspace*{0.5cm}
\epsfig{file=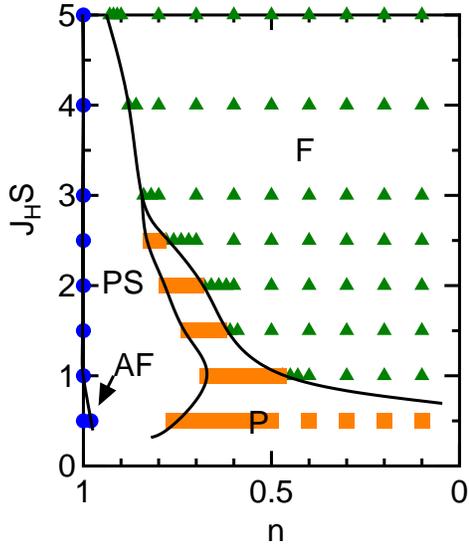}
  \label{sinfphase}
\end{figure}

\begin{figure}[hb]
  \caption{The magnetization for $S=\infty$, $\beta=400$ and $n=0.4$ as
  a function of $J_{\rm H}S$.}
\vspace*{0.5cm}
\epsfig{file=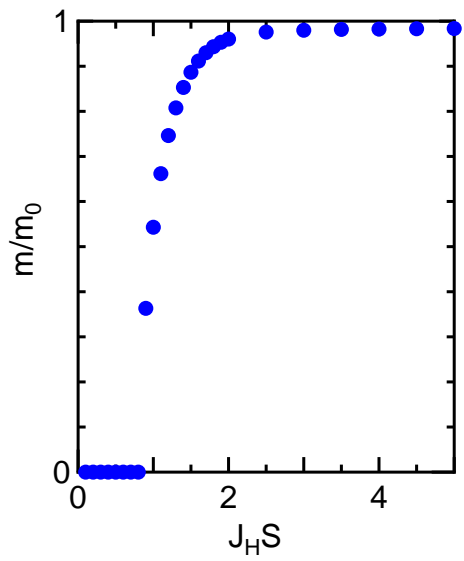}
  \label{sinfmag}
\end{figure}

\begin{figure}[ht]
  \caption{Wave-vector dependence of the spin susceptibility for 
    $J_{\mbox{{\scriptsize H}}}S=1.0$,
    $n=0.66$ and $\beta=110$.}
\vspace*{0.5cm}
\epsfig{file=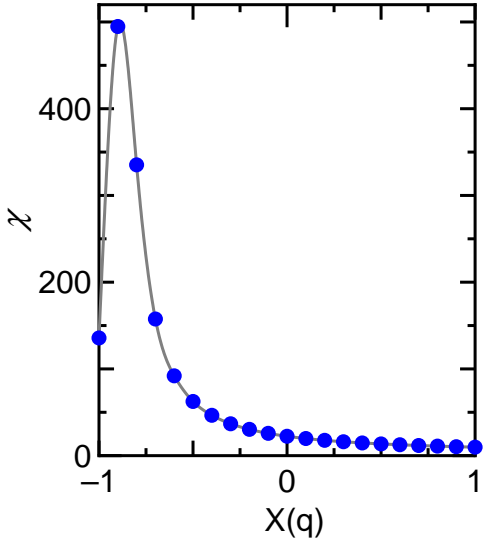}
  \label{Icnpara}
\end{figure}

\begin{figure}[hb]
  \caption{Phase diagrams of the ground state derived by the exact
    diagonalization 
    method with $N_{\mbox{{\scriptsize S}}}=6$ (a) and 
    $N_{\mbox{{\scriptsize S}}}=8$ (b). AF, F and P denote
    antiferromagnetic, ferromagnetic and paramagnetic phases, respectively.
    PS means the region where phase separation occurs.} 
\vspace*{0.5cm}
\epsfig{file=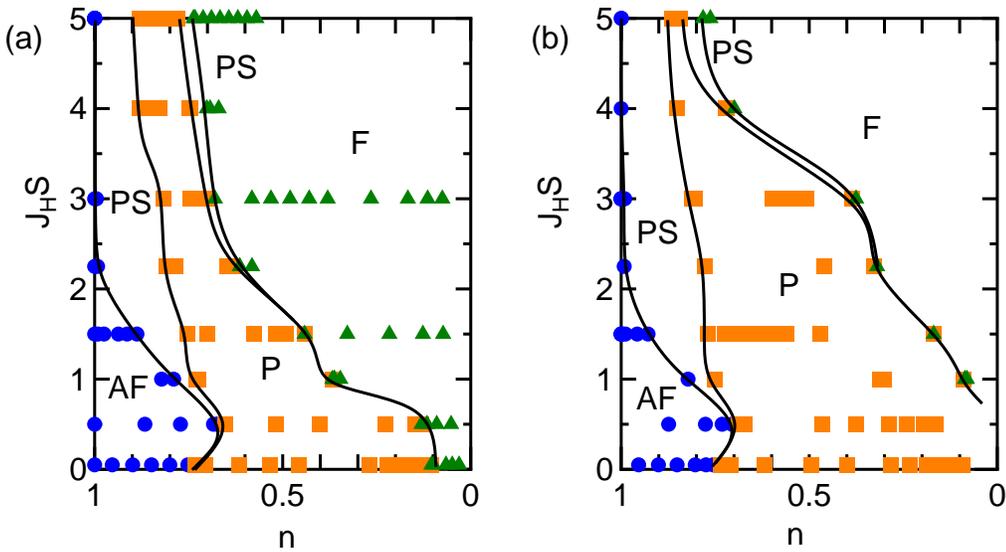}
  \label{phase}
\end{figure}

\begin{figure}[ht]
  \caption{The electron density $n$ as a function of $\mu$
    in the ground state
    for $J_{\mbox{{\scriptsize H}}}S=5.0$ and $N_{\mbox{{\scriptsize
          S}}}=8$. Circles, squares and triangles indicate
    the antiferromagnetic, paramagnetic and
    ferromagnetic states, respectively.} 
\vspace*{0.5cm}
\epsfig{file=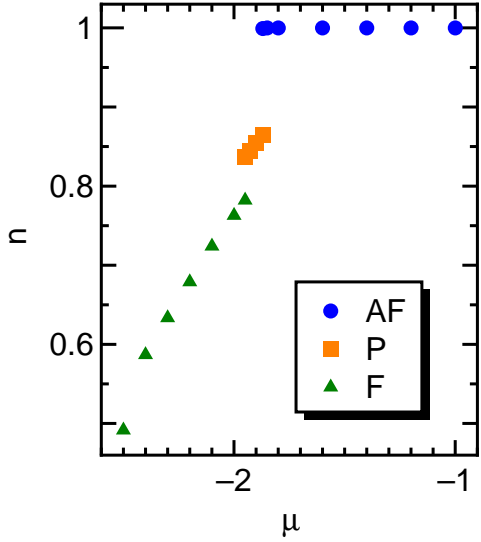}
  \label{munfig}
\end{figure}

\begin{figure}[hb]
  \caption{The kinetic energy (a) and the magnetization (b) of
the ground state as functions of 
    $J_{\mbox{{\scriptsize H}}}S$ for $n=0.4$ and 
    $N_{\mbox{{\scriptsize S}}}=8$. Rhombuses and squares indicate 
    the paramagnetic and ferromagnetic states, respectively.}
\vspace*{0.5cm}
\epsfig{file=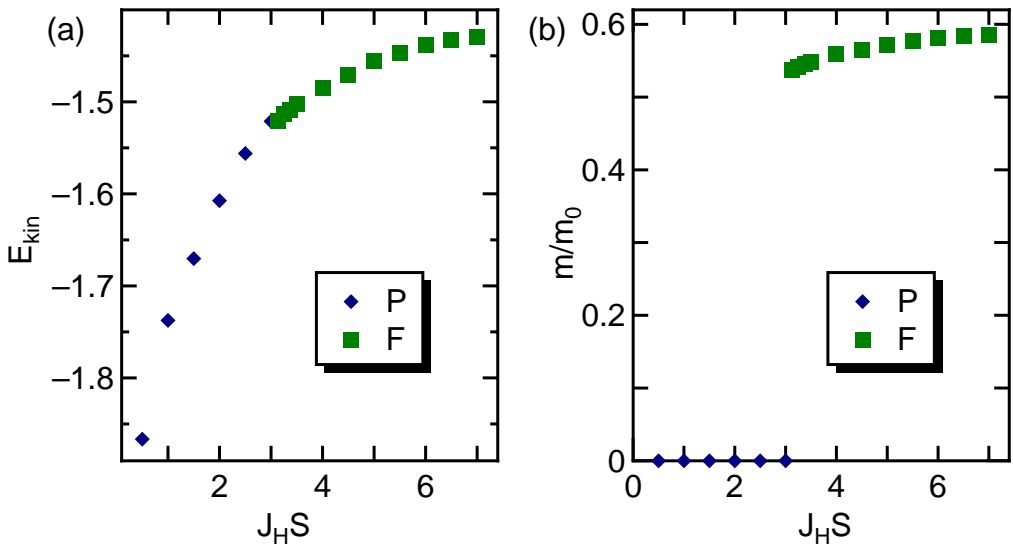}
  \label{kinene}
\end{figure}

\begin{figure}[ht]
  \caption{Density of states for $J_{\mbox{{\scriptsize H}}}S=5.0$ 
    at $n =1.0$ (a) and $n=0.78$ (b), and for 
    $J_{\mbox{{\scriptsize H}}}S=0.5$ at $n =1.0$ (c) and 
    $n =0.73$ (d).} 
\vspace*{0.5cm}
\epsfig{file=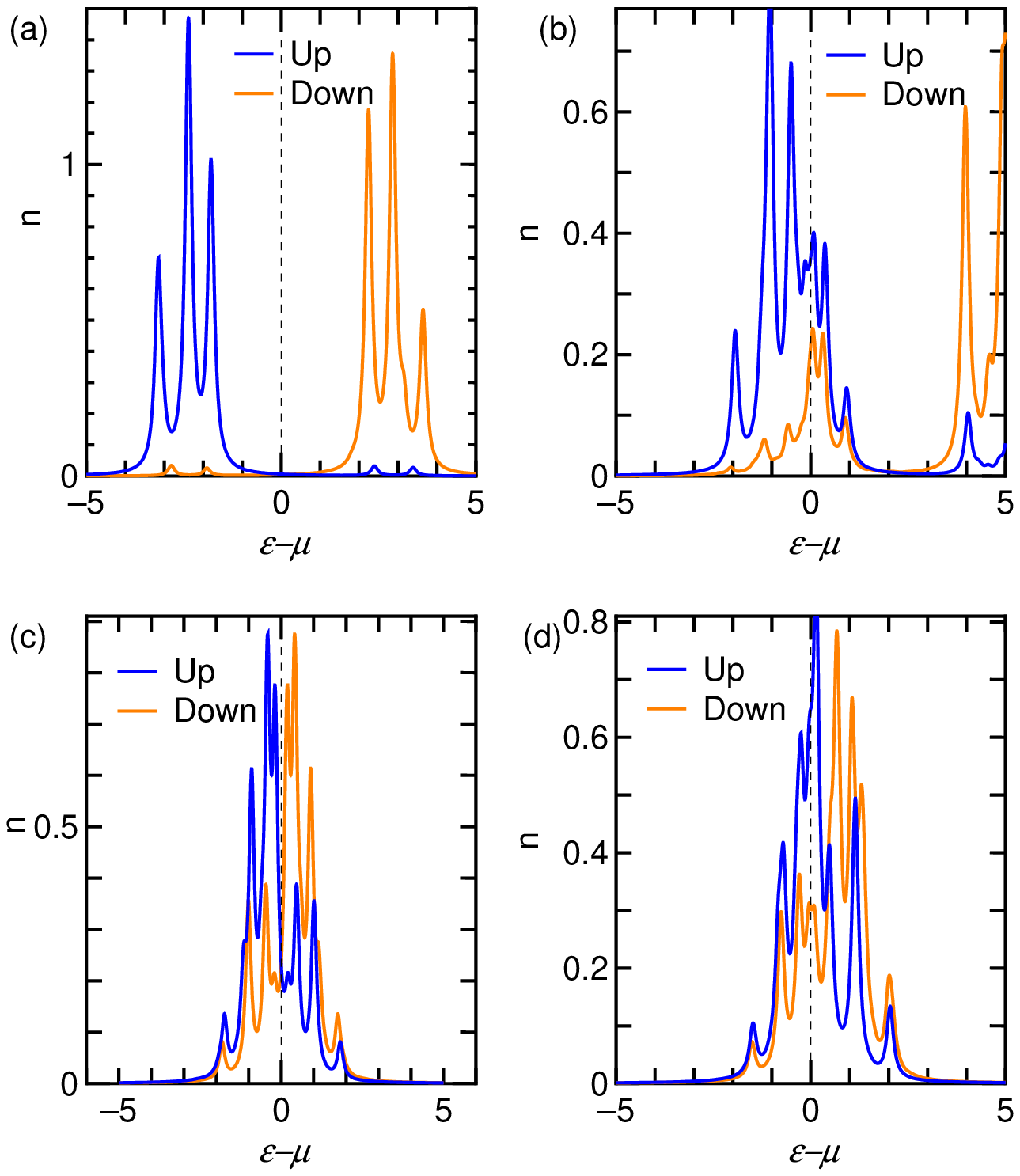}
  \label{dosfig}
\end{figure}

\end{document}

%% file: paper1.tex
\section{Introduction}
Recently perovskite manganese oxides  doped
with divalent alkaline metals, such as La$_{1-x}$Sr$_x$MnO$_3$ and
La$_{1-x}$Ca$_x$MnO$_3$, are studied extensively\cite{manganites}.
 In these materials
the decrease of 3$d$ electrons caused by the doping with divalent metals
leads to drastic changes of magnetic and transport properties. For
example, the mother material LaMnO$_3$ is insulating and has an
antiferromagnetic (AF) order at low temperatures. This AF insulator is 
transformed to a ferromagnetic (F) one by doping, and the
F state becomes metallic for $x\gsim 0.20$\cite{jonker}. 
The orthorhombic crystal structure changes into a rhombohedral one with
higher symmetry in the metallic phase. The so-called colossal negative
magneto-resistance shown by these materials is attracting intense
interests due to its potential applications\cite{colossal}.

Zener proposed the double exchange (DE) mechanism, which describes
essential physics in these materials\cite{zener}. In this paper we
study the simplest model that contains this mechanism, i.e. the DE
model. In this model three $t_{\mbox{{\scriptsize 2g}}}$ electrons on
a Mn ion are treated as a localized spin and the two-fold degeneracy of
$e_{\mbox{{\scriptsize g}}}$ orbitals is neglected. The Hamiltonian is
defined as 
\begin{equation}
H=-t\sum_{<i,j>,\sigma}c^{\dag}_{i\sigma}c_{j\sigma}-J_{\mbox{{\scriptsize
      H}}}\sum_{i} 
\bm{S}_i\cdot\bm{s}_i. 
\end{equation}
Here $t$ denotes the hopping amplitude between the nearest neighbor sites,
$c_{i\sigma}$ ($c^{\dag}_{i\sigma}$) the annihilation (creation)
operator of an $e_g$ electron at the site $i$ with spin $\sigma$, and
$\bm{S}_i$ and $\bm{s}_i$ are the localized spin and the spin of the
$e_{\mbox{{\scriptsize g}}}$ electron at the site $i$,
respectively. Itinerant $e_{\mbox{{\scriptsize g}}}$ electrons are assumed
to occupy a single orbital  and interact with the localized
 spins through the
Hund coupling $J_{\mbox{{\scriptsize H}}}$. 

Anderson and Hasegawa\cite{andhas} showed that in the limit of 
strong Hund coupling the hopping amplitude of electrons between the
sites $i$ and $j$ is reduced as $\tilde{t}_{ij}=t\cos (\theta_{ij}/2)$
 when the angle between the localized spins $\bm{S}_i$ and $\bm{S}_j$
is $\theta _{ij}$. The hopping amplitude is largest when
$\theta_{ij}=0$, which implies that the F state minimizes the
kinetic energy. This is the DE mechanism for the ferromagnetism 
proposed by Zener\cite{zener}. The competition between the DE mechanism
and the antiferromagnetic super-exchange interactions between the
localized spins was studied by de Gennes\cite{degen}. Kubo and Ohata
studied the DE model in the strong coupling limit\cite{kuboohata}. 

Recently Furukawa studied this model with classical localized spins 
using the dynamical mean field theory (DMFT), which is exact in
infinite dimensions\cite{furukawa0,furukawa1,furukawa2,furukawa3}. 
In DMFT the effective action for the system is
equivalent to that for a single impurity problem, and the momentum
dependence is absent in the self-energy of the one-particle Green
function. These features of the DMFT make the calculations of various
properties tractable. He obtained the thermodynamic as well as the
transport properties of the model employing a semicircular as well as
a Lorentzian density of states for non-interacting
electrons.
The DE model
with classical spins was studied also in finite dimensions with
numerical methods. Owing to these studies the properties of the DE
model with classical localized spins are fairly well understood. The
DE model with quantum localized spins is also studied. The spin wave
spectrum\cite{spinwave} and the ground state phase 
diagram\cite{riera,dagotto1} in one and two dimensions were studied 
by numerical
diagonalization. Variational arguments\cite{okabe,brunton} show that
the fully polarized state is destabilized in a certain density region
even in the strong coupling limit. 
Electronic states were also studied by a many-body CPA 
approximation\cite{edwards}. Due to the difficulty
of the strong correlation effects, we have still much to understand
the effects of quantum fluctuations in the DE model.    

In this paper, we study the DE model in the cases of $S=1/2$ and
the classical (the limit $S=\infty$ with finite $J_{\mbox{{\scriptsize
      H}}}S$) localized spins  in infinite dimensions. Comparing the 
magnetic phase
diagrams in both cases, we clarify the
influence of the quantum fluctuations of the localized spins on the
magnetic ordering. We use the exact diagonalization method in the
framework of the DMFT in order to study the case of $S=1/2$. We
employ the Gaussian density of states for non-interacting electrons, 
which is exact for the simple hypercubic lattice in infinite dimensions, 
\begin{equation}
D_{0}(\epsilon)=\frac{1}{\sqrt{2\pi}}\exp\kakkoa{-\frac{\epsilon^2}{2}}.
\label{doseq}
\end{equation}

The contents of this paper are as follows: In section 2 we summarize
the method of studying the DE model in infinite dimensions. In section 3
we report  the phase diagram and the magnetic susceptibility for the
case of  $S=\infty$. In section 4 we show the results for the case of
$S=1/2$ and compare them with those for $S=\infty$. In section 5 we
summarize the results and give perspectives for future study.

%% file: paper2.tex

\section{Method}

We employed the DMFT to study the DE model (1) in  infinite
dimensions. Though the method is now a standard
one\cite{DMFT}, we briefly
describe it in this section for readers who are not familiar with
 the method. In
infinite dimensions, the influence from neighboring  sites can be
replaced with a dynamical mean field (Weiss field). The problem is
hence reduced to
a single impurity problem in the Weiss field. The self-energy
of the Green function does not depend on the 
momentum\cite{metzner,mueller-hartmann}. The Green
function is written as 
\begin{equation}
G_{\sigma}(\bm{k},i\omega_n)=\frac{1}{i\omega_n-\epsilon_{\bm{k}}
-\mu-\Sigma_{\sigma}(i\omega_n)},
\end{equation}
where $\Sigma_{\sigma}(i\omega_n)$ is the self-energy and $\sigma$
denotes the spin suffix. It depends on the momentum $\bm{k}$ only
through $\epsilon_{\bm{k}}$. In the following text we omit the variables
$\bm{k}$, $i\omega$ and $\sigma$ in $G$, $\Sigma$ etc. for
simplicity.  The effective action of the DE model at a
temperature $T=\beta^{-1}$ is 
\begin{eqnarray}
{\cal S} & = & -\sum_{\sigma}\int^{\beta}_{0}d \tau_1\int^{\beta}_0d\tau_2
\Psi^*_{\sigma}(\tau_1){\cal G}_{0\sigma}^{-1}(\tau_1-\tau_2)
\Psi_{\sigma}(\tau_2) \nonumber \\
& & -J_{\mbox{{\scriptsize H}}}\int^{\beta}_{0}d
\tau\bm{S}(\tau)\cdot\bm{s}(\tau), 
\end{eqnarray}
where $\Psi^*_{\sigma}(\tau)$ and $\Psi_{\sigma}(\tau)$ are Grassmann
variables representing the electrons at the impurity site, 
$\bm{S}(\tau)$ the impurity localized spin, and ${\cal G}_0^{-1}$
is the Weiss function that represents a time-dependent mean
field. This Weiss function is related to the Green function $G$
and the self-energy $\Sigma$ as ${\cal G}_0^{-1}=G^{-1}-\Sigma$. 

We obtain the Weiss function ${\cal G}_0$ and the Green function $G$
self-consistently through the following procedures.    
\begin{enumerate}
\item Given ${\cal G}_0^{-1}$, then calculate  $G$
  employing the effective action. 
\item Calculate $\Sigma$ from $G$  through the
  relation ${\cal G}_0^{-1}=G^{-1}-\Sigma$. 
\item Calculate $G$ from $\Sigma$ as
\begin{equation}
G(i\omega _n)=\int d\epsilon \frac{D_{0}(\epsilon)}
{i\omega _n-\epsilon -\mu -\Sigma},
\end{equation}
where $D_{0}(\epsilon)$ is the density of states given by (\ref{doseq}).
\item Obtain ${\cal
    G}_0^{-1}$  again by using ${\cal
    G}_0^{-1}=G^{-1}-\Sigma$.
\end{enumerate}
We repeat the procedures from A to D until a 
self-consistent  solution is obtained. 

The main difficulty in these procedures lies in the procedure A, where we
must solve a quantum many-body problem. For finite $S$, we cannot
accomplish this procedure exactly and hence we need an
approximation.
The Weiss function is expressed as
\begin{eqnarray}
{\cal G}_{0\sigma}(i\omega_n)^{-1} & = & i\omega_n + \mu + 
\int_{-\infty}^{\infty}d\omega '\frac{\Delta_\sigma(\omega ')}
{i\omega_n-\omega '} \nonumber \\
& = & i\omega_n + \mu 
+ \sum _{p=2}^{\infty}\frac{V_{p\sigma}^2}{i\omega_n
-\tilde{\epsilon}_{p\sigma}},
\end{eqnarray}
where $\Delta_{\sigma}(\omega)=\sum_p^{\infty}V_{p\sigma}^2\delta(\omega-\tilde{\epsilon}_{p\sigma})$.
In this study we approximate ${\cal G}_{o\sigma}(i\omega_n)$ with
${\cal G}_{o\sigma}^{N_{\rm S}}(i\omega_n)$ which is the Weiss function
for a finite number of discrete orbitals;
\begin{equation}
{\cal G}_{o\sigma}^{N_{\rm S}}(i\omega_n)^{-1}= i\omega_n + \mu 
+ \sum _{p=2}^{N_{\rm S}}\frac{V_{p\sigma}^2}{i\omega_n
-\tilde{\epsilon}_{p\sigma}}.
\label{discretize}
\end{equation}
The parameters $V_{p\sigma}$ and
$\tilde{\epsilon}_{p\sigma}$ are determined self-consistently as to
minimize 
\begin{equation}
R = \frac{1}{n_{\mbox{\scriptsize max}}+1}
\sum^{n_{\mbox{\scriptsize max}}}_{n=0}|{\cal G}_0(i\omega_n)^{-1}
-{\cal G}^{(N_{\mbox{{\scriptsize S}}})}_0(i\omega_n)^{-1}|,
\end{equation}
where $n_{\mbox{\scriptsize max}}$ labels the maximum frequency 
employed in the calculation. When the Weiss function is approximated
as above, the problem is identical to solve the following impurity
Anderson Hamiltonian,
\begin{eqnarray}
{\cal H} & = & \sum_{p\geq 2,\sigma}^{N_{\mbox{{\scriptsize S}}}}
\tilde{\epsilon}_{p\sigma} a^+_{p\sigma}
a_{p\sigma}
+\sum_{p\geq 2,\sigma}^{N_{\mbox{{\scriptsize S}}}}
V_{p\sigma}(a^+_{p\sigma}d_{\sigma}+d^+_{\sigma}
a_{p\sigma}) \nonumber \\
& &+\epsilon_d\sum_{\sigma}d^+_{\sigma}d_{\sigma}-J_{\mbox{{\scriptsize
      H}}}\bm{S}\cdot\bm{s}, 
\end{eqnarray}
where $a^+_{p\sigma}$ ($a_{p\sigma}$) and $d^+_{\sigma}$ ($d_{\sigma}$)
are the creation (annihilation) operators of the orbital $p$ and 
the impurity site, respectively, and 
$s^{\alpha}=\frac{1}{2}\sum_{\sigma ,\sigma '}
d^+_{\sigma}\sigma^{\alpha}_{\sigma ,\sigma '}d_{\sigma '}$. We obtain  
the ground
state of this Hamiltonian with use of the exact diagonalization
method, and calculate the Green function. The kinetic and 
interaction energies in the ground state $\ket{\mbox{G.S.}}$ are
obtained with use of the expressions\cite{lanczos3}, 
\begin{equation}
E_{\rm kin}=\bra{\mbox{G.S.}}\sum_{p\geq 2,\sigma}^{N_{\mbox{{\scriptsize S}}}}
V_p(a^+_{p\sigma}d_{\sigma}+d^+_{\sigma}
a_{p\sigma})\ket{\mbox{G.S.}}
\end{equation}
and
\begin{equation}
E_{\rm int}=\bra{\mbox{G.S.}}-J_{\mbox{{\scriptsize
      H}}}\bm{S}\cdot\bm{s}\ket{\mbox{G.S.}}. 
\end{equation}

Although the size of a localized spin is
$3/2$ in actual manganese oxides, we adopt the case of $S=1/2$ in order to emphasize the quantum effects
as well as  for simplicity. The approximation improves with 
increase of the number $N_{\mbox{{\scriptsize S}}}-1$ of the assumed
orbitals. The present study was restricted to
$N_{\mbox{{\scriptsize S}}}
\le 8$ due to technical reasons. 

We also calculate the case of $S=\infty$ for the sake of
comparison. In this case the procedure A is easily accomplished
because the effective action is that of non-interacting electrons. The
partition function in the  case of $S=\infty$ is given by
\begin{equation}
Z  =  \int d\Omega_{\bm{S}}\int \prod_{\sigma}{\cal D}\Psi^*_{\sigma}
{\cal D}\Psi_{\sigma}\exp \kakkoa{-\cal{S}(\bm{S})}
\end{equation}
with 
\begin{eqnarray}
\cal{S}(\bm{S}) & = & \sum_{\sigma}\int^{\beta}_{0}d 
\tau_1\int^{\beta}_0d\tau_2
\Psi^*_{\sigma}(\tau_1){\cal G}_{0\sigma}^{-1}(\tau_1-\tau_2)
\Psi_{\sigma}(\tau_2) \nonumber \\
& & -J_{\mbox{{\scriptsize H}}}\int^{\beta}_{0}d \tau\bm{S}\cdot\bm{s}(\tau),
\end{eqnarray}
where $\bm{S}$ is a classical vector of size $S$ and 
$\Omega_{\bm{S}}$ denotes the volume angle of the direction of the
localized spin.
The Green function is given by
\begin{equation}
G_{\sigma}(i\omega_n)=\frac{1}{Z}\int d\Omega_{\bm{S}}
P(\bm{S})\kakkoa{{\cal G}_{0\sigma}^{-1}
(i\omega_n)+J_{\mbox{{\scriptsize H}}}\bm{S}\cdot\bm{s}(i\omega_n)}^{-1},
\end{equation}
where $P(\bm{S})$ is the Boltzmann weight for the configuration of the
localized spin, 
\begin{equation}
P(\bm{S})=\int \prod_{\sigma}{\cal D}\Psi^*_{\sigma}
{\cal D}\Psi_{\sigma}\exp \kakkoa{-{\cal S}(\bm{S})}.
\end{equation}

%% file: paper3.tex
\section{Results for $S=\infty$}
In this section we report the temperature dependence of the magnetic
susceptibility and the phase diagram at a very low temperature in the
case of $S=\infty$. 

In infinite dimensions the susceptibility depends on the wave vector
$\bm{q}$ only
through $X(\bm{q})$\cite{mueller-hartmann}, where $X(\bm{q})$ is defined by 
\begin{equation}
X(\bm{q})=\frac{1}{d}\sum_{i=1}^d \cos q_i.
\label{Xdef}
\end{equation}
Here $d$ denotes dimensionality of
the space. The F and AF
susceptibilities correspond to the cases with $X(\bm{q})=1.0$ and
$-1.0$, respectively, and  a
case where $-1.0 < X(\bm{q}) < 1.0$  corresponds to an incommensurate order.
 Figure \ref{J10susa} shows the magnetic
susceptibility $\chi$ for $J_{\mbox{{\scriptsize H}}}S=5.0$ at the
electron density $n=1.0$. Figure \ref{J10susa}(a) exhibits the
wave-vector dependence of the susceptibility at various temperatures. 
The
susceptibility has maximum at $X(\bm{q})=-1.0$ at all temperatures and
it increases with decrease of temperature. In Fig.\ 
\ref{J10susa}(b) the inverse of the AF susceptibility
is shown as a function of temperature. The susceptibility obeys the
Curie-Weiss law and diverges at a finite temperature
$T_{\mbox{{\scriptsize N}}}$. As a result  the AF order
is expected to appear at 
temperatures below $T_{\mbox{{\scriptsize N}}}\cong 0.044$ 
for $J_{\mbox{{\scriptsize H}}}S=5.0$ and
$n=1.0$. Next, the
susceptibility for $J_{\mbox{{\scriptsize H}}}S=5.0$ and $n=0.8$ is
shown in Fig.\ \ref{J10susf}. In this case the F
susceptibility is largest, and obeys the Curie-Weiss law. The
F order is expected to appear at temperatures below
$T_{\mbox{{\scriptsize C}}}\cong 0.062$.

We obtained a phase diagram at a very low temperature ($\beta=400$)
assuming the uniform or a two sublattice structure. The result is
shown in Fig.\ \ref{sinfphase}, where the AF, 
F and paramagnetic (P) phases are present. 
For $n=1$ the AF state is stable for any  positive $J_{\mbox{{\scriptsize
      H}}}$. The AF phase spreads out to a small 
region close to the line $n=1$  for weak Hund coupling. 

The F phase occupies a large region where $n\lsim 0.5$ and
$J_{\scriptsize H}S\gsim 1$. This F state has almost fully
polarized magnetization.
In Fig.\ \ref{sinfmag}, we show the spontaneous magnetization for
$n=0.4$ at $\beta=400$ as a function of $J_{\rm H}S$. The
magnetization starts increasing continuously at $J_{\rm H}S\cong 0.9$
and rapidly approaches the limiting value of $0.98m_0$ where $m_0$ is
the value of the full polarization. In this case the transition to
the F state is of the second order. 
The P phase appears for 
$J_{\scriptsize H}S\lsim 3$ and it spreads to a wide density region for 
weak coupling.  Phase separation (PS) occurs between the AF and P
phases for $J_{\mbox{{\scriptsize H}}}S\lsim 3$ and also between the
AF and F phases for $J_{\mbox{{\scriptsize H}}}S\gsim 3$.
General features of the phase diagram is similar to those obtained previously 
in infinite dimensions\cite{yunoki1,dagotto1} and roughly understood as a
result of the competition between the AF coupling, which is  a  second
order effect of the hopping,  and the ferromagnetic DE 
mechanism. Since the AF coupling 
decreases with increasing $J_{\mbox{{\scriptsize H}}}S$ as
 $t^{2}/J_{\mbox{{\scriptsize H}}}S$, the relative strength of the F
 coupling, which is proportional to $t$,  increases with
 $J_{\mbox{{\scriptsize H}}}$.  
Occurrence of the PS for small doping ($n\simeq 1$) may be understood
 as the instability of a canted AF state\cite{golosov}.  

Since we assumed a two-sublattice structure there remains the
possibility that we missed magnetic phases with more complicated
spatial structures, e.g.  
those  with  incommensurate (IC) wave vectors. Such IC order was predicted 
previously to occur for intermediate
doping\cite{inoue,yunoki1}. The obtained P phase might be an IC
phase in reality.  In fact we found the cases where  
$\chi$ corresponding to an IC order diverges at higher temperatures. 
Figure 4 shows $\chi$ for $J_{\mbox{{\scriptsize H}}}S=1.0$, $n=0.66$
and $\beta =110$. 
We observe a sharp maximum of $\chi$ at $X(\bm{q})\simeq -0.9$ and it
diverges at a critical  temperature. It is certain that an IC phase
exists below the critical temperature. Study of the phase diagram
which contains IC phases is left for a future study.
%


%% file: paper4.tex
\section{Results for  $S=1/2$}
Quantum Monte Carlo method is usually used in the framework of the
DMFT for studying the thermodynamic properties at finite
temperatures\cite{DMFT}. This method, however, suffers from a
serious negative-sign problem in the study of the DE model with 
quantum localized spins\cite{nagai}.
In this section, we therefore studied only the ground state properties
for the model 
with $S=1/2$ localized spins, employing the exact diagonalization 
technique. We report the results and
compare them with those for the case of $S=\infty$. 

Figures \ref{phase}(a) and \ref{phase}(b) show the ground state phase 
diagram obtained with $N_{\rm S}=6$ and 8, respectively. We should
extrapolate the results of finite $N_{\rm S}$ to the limit 
$N_{\rm S} = \infty$ in order to obtain the exact result.
We observe that the two phase diagrams are almost same except for the 
boundary of the F phase. 
Difference between results for $N_{\rm S}=6$ and 8
is still large on the boundary of the F phase and hence the obtained 
boundary may not represent that of 
$N_{\rm S} = \infty$ correctly. Nevertheless 
these results exhibit contrast to that for the case of $S=\infty$ in 
several points.

The  AF phase appears in a region close to the line $n=1$ and its
boundaries for  $N_{\rm S}=6$ and 8 agree very well. It is remarkable
that the AF phase is much more stabilized for $S=1/2$  than  for
$S=\infty$ 
in the weak coupling region. It extends to $n\simeq 0.7$ and, as a
result, the region of PS narrows for small  $J_{\mbox{{\scriptsize
H}}}$ in contrast to the result for $S=\infty$. The present result 
shows that the AF phase is stable for $0.8\lsim n\le1$ 
even for $J_{\mbox{{\scriptsize H}}}=0.1$, the smallest 
 $J_{\mbox{{\scriptsize H}}}$ studied, though  the system should be
paramagnetic for all $0<n<1$ for  $J_{\mbox{{\scriptsize H}}}=0$. 

The F phase appears for 
 $J_{\mbox{{\scriptsize H}}}S\gsim 4$ in a wide range of the density, 
i.e.,  $n\lsim 0.7 $. Results for $N_{\rm S}=6$ and 8 agree well in 
this region. On the other hand, the F phase is strongly reduced for 
 $J_{\mbox{{\scriptsize H}}}S\lsim 3$ by increasing  $N_{\rm S}$
and almost disappears for  $J_{\mbox{{\scriptsize H}}}S\lsim 1$
in Fig.\ \ref{phase}(b). In comparison with the $S=\infty$ case, the F 
phase region for $N_{\rm S}=8$ is fairly reduced. We may need further 
study with larger $N_{\rm S}$  
to confirm this strong reduction of the F phase. 
 
Between the AF and F phases we obtain a  P phase in a wide density range,
and PS occurs between the AF and F phases as well as between the P
and F phases. The P phase persists to large  $J_{\mbox{{\scriptsize
H}}}$, i.e. 
 $J_{\mbox{{\scriptsize H}}}S\simeq 5$ in contrast to the case of
$S=\infty$ where the P phase terminates at 
$J_{\mbox{{\scriptsize H}}}S\simeq 3$.
We should note that  the obtained P phase might be replaced by phases
with more complicated spatial structures, e.g. IC phases, in more
comprehensive calculations, which include larger-sublattice structures. 

Figure 6 shows the electron density $n$ as a function of the chemical 
potential $\mu$ for 
$J_{\mbox{{\scriptsize H}}}S=5.0$ and $N_{\rm S}=8$. We see jumps of $n$
at two values of $\mu$, which indicate the occurrence of PS's. The
density region of the PS between the AF and P phases narrows with
decreasing $J_{\mbox{{\scriptsize H}}}$ 
in contrast to the case of $S=\infty$. This is caused by the expansion of 
the AF phase in the weak coupling region. The PS occurs in a very narrow 
density region between the P and F phases for
$J_{\mbox{{\scriptsize H}}}S\gsim 2$. This PS was not found in the
case of $S=\infty$. 

In Figs.\ 7(a) and 7(b), we show the kinetic energy and the
magnetization of conduction electrons, respectively, as functions  of 
$J_{\mbox{{\scriptsize H}}}S$ for $n=0.4$.
The kinetic energy [Fig.\ \ref{kinene}(a)] increases with 
$J_{\mbox{{\scriptsize H}}}$ very rapidly in the P phase, while it
shows a saturation in the F phase for $J_{\mbox{{\scriptsize H}}}S \ge
3.5$. This behavior is a characteristic of the ferromagnetism caused
by the DE mechanism and indicates that the scattering by the Hund
coupling ceases to be operative by alignment of the 
localized spins\cite{kubo}. 
In Fig.\ \ref{kinene}(b), the F phase starts with a finite
magnetization at $J_{\mbox{{\scriptsize H}}}S = 3.5$, which indicates
a first-order transition from the P phase. The occurrence 
of PS between the F and P phases is related to the first-order
character of the phase transition. 
The magnetization increases
with $J_{\mbox{{\scriptsize H}}}$ and apparently approaches to a finite value 
nearly equal to 0.6$m_{0}$, 
where $m_0$ denotes the magnetization of a fully polarized state. 
It is noticeable that 
the saturated moment is relatively small in comparison with the result 
for the $S=\infty$ localized spins. This result gives 
evidence that the quantum fluctuations of the localized spins 
destabilize ferromagnetism. Though the fully polarized state 
cannot be realized for the Gaussian density of states, the reduction 
of the magnetization mostly comes from the quantum fluctuations of 
the localized spins.

We also studied the electronic states in the ground states. 
The local density of states (DOS) is calculated from the Green function
obtained in the procedure A in section 2.
Because we treat the system with a finite number of discrete orbitals,
the obtained DOS is composed of finite number of sharp peaks.
Figures \ref{dosfig}(a-d) exhibit examples of  the DOS obtained by
calculations with $N_{\mbox{{\scriptsize S}}}=8$, where an artificial 
imaginary part $0.1i$ is added to $\Sigma(\omega)$ for the purpose of
illustration. Figure \ref{dosfig}(a) exhibits the local DOS in an AF
ground state for $J_{\mbox{{\scriptsize H}}}S=5.0$ at $n=1.0$. 
We can observe characteristic features of the DOS in the AF ground state
in the strong coupling region (in spite of the discretized structures
due to the approximation). In this case DOS is composed of two
sub-bands each of which 
corresponds to electronic states with parallel and antiparallel spins
with localized spins. Two sub-bands are separated from
the Fermi level by a broad energy gap, which indicates that the ground 
state is insulating. 
The lower DOS for up spin has a large weight compared to that for
down spin, since spin is almost fully polarized upwards 
on this site.  
Figure \ref{dosfig}(b) shows the local DOS of a F ground state for 
$J_{\mbox{{\scriptsize H}}}S=5.0$ at $n=0.78$. 
Also in this case the DOS is composed of two sub-bands separated by an
energy gap (the main part of the upper sub-band is out of range of
the figure). The Fermi level is inside of the lower sub-bands, which 
indicates the metallic partially 
polarized ground state. The DOS for down spin is much smaller than
that for 
up spin, which implies that the spins are  polarized upwards. 
The width of the lower sub-bands for both spins are almost same, 
which reflects the mixing due to exchange scattering. 
The DOS in the AF ground state for $J_{\mbox{{\scriptsize H}}}S=0.5$ at 
$n=1.0$ is exhibited in Fig.\ \ref{dosfig}(c). 
In this case the DOS is apparently  composed of a 
single band centered around the Fermi level. It is 
expected that there is an energy gap caused by the AF order 
and/or Umklapp scatterings. However it is exhibited as a dip at the
Fermi level. This is only due to the width added 
artificially. The DOS for down spin  below Fermi level is smaller 
than, but comparable to that for up
spin due to small sublattice magnetization. We show in Fig.\
\ref{dosfig}(d) the DOS for  $J_{\mbox{{\scriptsize H}}}S=0.5$ and 
$n=0.73$ where the ground state is antiferromagnetic. 
In this case the dip located at $\epsilon = \mu + 0.3$ may correspond
to the gap due to the AF ordering. The Fermi level lies very close to
the top of a peak. These results suggest that the ground state is
a metallic state with AF long range order.

%% file: paper5.tex

\section{Summary and future work}

We studied the DE model with $S=1/2$  as well as $S=\infty$ localized
spins in infinite dimensions with use of the DMFT. 
It is found that quantum fluctuations of localized spins partly 
destabilize the ferromagnetic ground state and the paramagnetic phase 
expands in the case of $S=1/2$. 
The F state for $S=1/2$ has rather small polarization 
even for a strong coupling, such as  $J_{\mbox{{\scriptsize H}}}S\gsim 5.0$. 
This is in contrast to the results for $S=\infty$, where the ground
states were almost fully polarized\cite{yunoki1,dagotto1}. 
On the other hand the AF ground state is stabilized by quantum effects 
for weak coupling. 
A similar result was recently reported for a model with degenerate 
orbitals \cite{chatto}.
It is interesting to examine whether these results still hold for 
different choice of
DOS (for non-interacting electrons) from the Gaussian one. 

More accurate calculations to determine  the precise ground state 
phase diagram  is left for a future study. 
The obtained phase diagrams for the $S=1/2$ localized spin systems do
not show good convergence at the boundary of the ferromagnetic phase. 
Furthermore the  ground state was obtained under the assumption of
the two-sublattice structure.  We found that the phase transition to an
IC state occurs in the case of $S=\infty$ and we expect it to occur 
for $S=1/2$ as well. 
 
It is interesting to study  the case of $S=3/2$  
and compare the result with those for $S=1/2$ and $\infty$, 
since the  magnitude of the localized spins are 3/2 in real 
manganites. This is now under progress and will be reported 
elsewhere.

%% file: acknowledge.tex
\section{Acknowledgements}
The authors are indebted to  D.M.\ Edwards, A.C.M.\ Green, T.A.\ Kaplan,
N.\ Furukawa, D.S.\ Hirashima and H.\ Tsunetsugu for valuable 
discussions and helpful comments. 
This work  was supported by Grant in Aid Nos. 09640453 and 11640365 
from the Ministry of Education, Science and Culture of Japan. 
Numerical calculations were done on Facom VPP500 at the ISSP of the
University of Tokyo.